\documentclass[aps,prl,twocolumn,showpacs,preprintnumbers,amsmath,amssymb]{revtex4}

\newcommand{\nn}{\nonumber}

\newcommand{\be}{\begin{equation}}
\newcommand{\ee}{\end{equation}}
\newcommand{\bea}{\begin{eqnarray}}
\newcommand{\eea}{\end{eqnarray}}

\usepackage{graphicx}
\usepackage{dcolumn}
\usepackage{bm}
\usepackage{color}

\newcommand{\e}{\varepsilon}

\newcommand{\s}{\sigma}

\begin{document}

\title{Theory of frequency-dependent spin current noise through correlated quantum dots}

\author{C. P. Moca$^{1}$, I. Weymann$^{2,3}$, G. Zar\' and$^{1}$}
\affiliation{$^{1}$ Department of Theoretical Physics, 
Budapest University of Technology and Economics, H-1521
Budapest, Hungary\\
$^{2}$ 
Physics Department, 
Ludwig-Maximilians-Universit\" at, Theresienstrasse 37, 80333 Munich, Germany
\\
$^{3}$ 
Department of Physics, Adam Mickiewicz University, 61-614 Pozna\'n, Poland
}
\date{\today}

\begin{abstract}
We analyze the equilibrium and non-equilibrium frequency-dependent
spin current noise and spin conductance through a quantum dot in the 
local moment regime. Spin current correlations are shown to behave markedly 
differently from charge correlations: Equilibrium spin cross-correlations 
are suppressed at frequencies below the Kondo scale, and 
are characterized by a universal function that we determine numerically 
for $T=0$ temperature.  For asymmetrical quantum dots 
 dynamical spin accumulation resonance is found at the Kondo energy,
$\omega\sim T_K$. At higher temperatures
surprising low-frequency anomalies  related to overall spin conservation 
appear.
\end{abstract}

\maketitle


{\bf \emph{Introduction.}} Coherent detection and manipulation of
spin currents in nanostructures has recently attracted wide
attention due to possible promising applications in future storage
technologies and quantum computing~\cite{loss02, zutic04}. Many
proposals have been made to build spin batteries to inject
spin-polarized current, and then filter, manipulate, and detect
it~\cite{FolkarXiv08}. Often one makes use of  ferromagnetic
electrodes in these circuits~\cite{Sih05}, while in other cases
the application of an external magnetic field \cite{Potok02} or
the presence of a ferromagnetic resonance process
\cite{WatsonPRL03} enables one to filter and detect spin currents.
Quantum dots play a special and important role in this regard: In
these devices, the strong electron-electron interaction enables
one to manipulate the spin of a single electron~\cite{Sasaki2000},
and such quantum-dot devices provide a possible route to quantum
computing~\cite{LossPRA98}.

However, to use spin circuits efficiently,  it would be of crucial
importance to characterize the noise in them. In addition, the
structure of the noise provides valuable information on
interactions and correlations. In fact, a lot of attention has
been devoted to  noise analysis in correlated mesoscopic circuits
for this reason~\cite{KaminskiPRB00, MeirPRL2002, SindelPRL2005}.
Due to progress in experimental technology, it is now  possible to
measure {\it ac} conductance properties as well as
frequency-dependent noise in these circuits down to very low
temperatures, and  even in the Kondo regime~\cite{GabelliPRL08,
BillangeonPRL07, DelattreNatPhys09}. Furthermore, with efficient
spin filtering methods~\cite{FolkarXiv08} measuring spin-resolved
current noise in such circuits is also within reach. Surprisingly,
while a lot is known about the properties of ordinary noise in the
correlated regime, much less is known about the structure of spin
current noise. So far, only spin correlations in the sequential
tunneling~\cite{SauretPRL2004,CottetPRL2004} and perturbative
regimes~\cite{KindermannPRB2005} have been analyzed, and these
works focused almost exclusively on shot noise.

Here  we carry out a detailed analysis of the full frequency
spectrum of the spin-dependent current noise in the Kondo regime.
We show that equilibrium spin current correlations are
characterized by two universal functions, which we determine
numerically for $T=0$ temperature using the method of numerical
renormalization group (NRG)~\cite{WilsonRMP75}, and compute
analytically for large frequencies. At finite temperatures, we
analyze spin  correlations using a  perturbative approach. We find
in all regimes that correlations between electrons of the same and
opposite spins behave markedly differently.  In the perturbative
regime these remarkable differences emerge at frequencies below
the Korringa relaxation rate: while a dip appears in the
frequency-dependent noise of opposite spin directions, a large
peak develops for the parallel spin components. These surprising 
features are all intimately related to spin conservation.


{\bf \emph{ Model.}} Focusing on the Kondo regime, we shall assume
that there is a single spin $S=1/2$ electron on the quantum dot,
which couples to the electrons on the leads through the Kondo
interaction~\cite{GlazmanReview},
\begin{equation}
H_{\rm int}=  \sum_{r,r'=L,R} \sum_{\sigma,\sigma'}
\frac j 2 \; v_r v_{r'}\;
\mathbf{S} \;\psi_{r\sigma}^\dagger {\mathbf{\sigma}}_{\sigma\sigma'}
\psi_{r'\sigma'}\;.
\label{eq:H_int}
\end{equation}
Here $\mathbf \sigma$ stands for the three Pauli matrices, 
the fields $\psi_{r\sigma} = \int_{-D}^D  c_{r\s}(\e) d\e$
destroy electrons of spin $\sigma$ in leads $r\in\{L,R\}$, and
their dynamics are governed by the non-interacting Hamiltonian,
$H_0 = \sum_{r\s} \int \e\; c^{\dagger}_{r\s}(\e) c_{r\s}(\e) d\e$
\footnote{The annihilation operators $c_{r\s}(\e)$ satisfy
 $\{c^{\dagger}_{r\s}(\e),c_{r'\s'}(\e')\}
=\delta_{rr'}\delta_{\s\s'}\delta(\e-\e')$.}. The coupling $j$ in
Eq.~\eqref{eq:H_int}  is the usual dimensionless coupling, which
incorporates already the density of states in the leads, and is
related to the Kondo temperature as $T_K\approx D\; e^{-1/j}$,
with $D$ the cut-off energy appearing in $\psi_{r\sigma}$. The
dimensionless hybridization parameters  are given by
 $v_L = \cos(\phi/2)$ and $v_R = \sin(\phi/2)$, with
$\phi$  parametrizing the asymmetry of the dot: $\phi = \pi/2$
corresponds to a symmetrical quantum dot with maximum
transmittance.

{\bf \emph {Equilibrium noise.}} In view of the special structure
of Eq.~\eqref{eq:H_int}, it is natural to introduce the 'even' and
'odd' linear combinations, $\Psi \equiv \cos(\phi/2)\;\psi_L +
\sin(\phi/2)\;\psi_R $, and $\tilde \Psi \equiv
\sin(\phi/2)\;\psi_L - \cos(\phi/2)\;\psi_R $. Although only
$\Psi$ couples to the spin in $H_{\rm int}$, changing the chemical
potential in one of the leads couples the fields  $\tilde \Psi$
and $\Psi$, and both contribute to the spin noise.

To compute the noise, we first define the spin component $\s$ of
the current in lead $r$ through the equation of motion,
$J_{r\sigma}\equiv e\;\dot N_{r \sigma} = e\;i[H_{\rm int},N_{r
\sigma}]$. The corresponding current is found to have two distinct
(even and odd) parts, $J_{r\sigma}= I_{r\sigma} + \tilde
I_{r\sigma}$, with
\bea
I_{r\sigma} = e\;j \;\gamma_{r}\; i(F^\dagger_\sigma
\Psi_\sigma - \Psi_\sigma^\dagger  F_\sigma) \;, \nn
\\
\tilde I_{r\sigma} =  e\;j \;\tilde \gamma_{r} \; i
(F^\dagger_\sigma \tilde \Psi_\sigma -\tilde {\Psi}_\sigma^\dagger
F_\sigma)\;, \label{eq:evenodd}
\eea
and the prefactors defined as  $\gamma_{L/R} = [1 \pm \cos(\phi)]/4$ and
$\tilde \gamma_{L/R} = \pm \sin(\phi)/4$. The operator $F_\sigma =
(\mathbf{S {\mathbf \sigma}}\Psi)_\sigma$ denotes the so-called composite
fermion operator~\cite{CostiPRL94}, and represents the universal (Kondo)
part of the dot-electron.

The operator identity, $I_{r\uparrow} + I_{r\downarrow} =0$, and
the simple even-odd decomposition of $J_{r\sigma}$ imply that, in
equilibrium and in the absence of external magnetic field, the
sixteen components of the symmetrized noise
$S_{rr'}^{\sigma\sigma'} \equiv \frac 12 \langle \{
J_{r\sigma}(t), J_{r'\sigma'}(0) \}\rangle $, depend on just two
universal functions, $s$ and $\tilde s$. Maybe the most
interesting  left-right noise component, $S_{LR}^{\sigma\sigma'}$,
can be expressed, e.g., as
\be
S_{LR}^{\sigma\sigma'}(\omega) = -\frac {e^2} {2\pi} \;T_K\;
\sin^2(\phi) \; \left(\delta_{\sigma\sigma'} \tilde s  (\omega) +
{\sigma\sigma'}\;s(\omega)  \right)\;,
\nonumber
\ee
where $e^2/2\pi = e^2/h$ denotes the universal conductance unit,
and the dimensionless functions $s$ and $\tilde s$ depend
exclusively on the ratios $\omega/T_K$ and $T/T_K$. The function
$s$ is related to the 'even' current component, and it  governs
the correlations between spin up and spin down carriers, however,
its contribution cancels in the charge noise and charge
conductance, which are exclusively determined by the 'odd'
component of the current, incorporated in $\tilde s$.

%
%

In equilibrium, the fluctuation-dissipation theorem relates
$S_{rr'}^{\sigma\sigma'}(\omega)$ to the real part of the
spin-conductance through the dot, ${\rm
Re}\;G_{rr'}^{\sigma\sigma'}(\omega) = - \frac1 \omega \; {\rm
coth}(\omega/ 2 T)\; S_{rr'}^{\sigma\sigma'}(\omega)\;, $ which
can therefore also be expressed in terms of two dimensionless
universal conductance functions, $g  (\omega,T)$ and $\tilde g
(\omega,T)$. The left-right conductance, e.g., reads
$$
{\rm Re}\;G_{LR}^{\sigma\sigma'}(\omega) = \frac{e^2}{2\pi} \,\sin^2(\phi) \,
\bigl(\delta_{\sigma\sigma'} \tilde g  (\omega,T) +
{\sigma\sigma'}\;g(\omega,T)  \bigr )\;.
$$

Using tedious but straightforward manipulations, we can express
$g$ and $\tilde g$ in terms of the spectral functions
$\varrho_F(\omega,T)$ and $\varrho_{{\cal I}_\sigma\;{\cal
I}_{\sigma'}}(\omega,T)$ of  the composite fermion and of the
``current'' operator ${\cal I}_\sigma \equiv i(F^\dagger_\sigma
\Psi_\sigma - \Psi_\sigma^\dagger  F_\sigma)$,
\bea
\tilde g (\omega,T) &=& \frac 1 {2\omega}\int d\omega' \frac
{\varrho_F(\omega',T)}{\varrho_F(0)} \large[f(\omega' -\omega)-
f(\omega' + \omega)  \large]\;, \nn
\\
g(\omega,T) &=& - \frac 1  {2\omega\; \varrho_F(0)}\;
\varrho_{{\cal I}_\uparrow {\cal I}_{\uparrow}}(\omega,T) \;,
\eea
with $f(\omega)$ denoting the Fermi function \cite{longPRB}. Since
$F_\sigma$ and ${\cal I}_\sigma$ are local operators, we can
compute $g$ and $\tilde g$ (and thus $s$ and $\tilde s$) by using
the powerful method of numerical renormalization group (NRG)
\cite{WilsonRMP75,BudapestNRG}.

\begin{figure}
\includegraphics[width=0.8\columnwidth,clip]{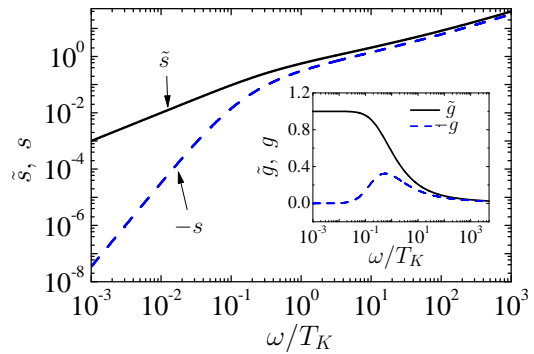}
\caption{\label{fig:universal_functions} (color online)
Zero-temperature universal functions $s$ and $\tilde s$ computed
by NRG. Inset: universal spin conductance functions $g$ and
$\tilde g$.}
\end{figure}


{\bf\em $T=0$, equilibrium results.} The $T=0$ temperature
universal functions $\tilde s(\omega/T_K)$, and $s(\omega/T_K)$
and the conductance functions $g(\omega/T_K)$ and $\tilde
g(\omega/T_K)$ are displayed in
Fig.~\ref{fig:universal_functions}. The high-frequency behavior of
$s$ and $\tilde s$ can be captured by doing perturbation theory in
$j$ and summing up the leading logarithmic corrections to give
\bea
&\tilde s(\omega/T_K) \approx -\frac 32 \; s(\omega/T_K) 
 \approx \frac{3\pi^2}{16} \frac{|\omega|}{T_K}
\frac{1}{\ln^2(|\omega|/T_K)}\;
\nn 
\eea
for $\omega \gg T_K$. Though they look similar at high
frequencies, $s$ and $\tilde s$ behave markedly differently in the
Fermi liquid regime, $\omega\ll T_K$, where $\tilde s = \tilde
\alpha |\omega|/T_K+\dots$, while $s =  \alpha \; (|\omega|/T_K)^3
+\dots$, with $\alpha$ and $\tilde\alpha$ universal constants of
the order of unity. The $\omega^3$ scaling of $s$ is related to
spin conservation: In the absence of external spin relaxation
mechanism, the total number of spin up electrons can fluctuate
between two values, $N_\uparrow$ and $N_\uparrow+1$. Since the
spin up electrons couple to the spin down electrons only at a
single point (the quantum dot), no steady spin current can be
generated for the spin down electrons by  injecting spin up
electrons in one of the leads. Thus the  spin conductance
$G_{rr'}^{\uparrow\downarrow}(\omega)$ must vanish at $\omega=0$,
and by analyticity, $G_{rr'}^{\uparrow\downarrow}\sim \omega^2$.
In equilibrium, however, the spin current noise is simply related
to the spin conductance by the fluctuation-dissipation theorem,
implying a $|\omega|^3$ scaling of $s$ at ${T=0}$. This argument
carries over to finite temperatures too,  where it leads to an
asymptotic behavior, $s\sim T \,\omega^2$ in the absence
of external spin relaxation. We should emphasize that, in our
calculations, spin relaxation is due to the interaction part of
the Hamiltonian which, however, conserves the total spin, and
leads to the vanishing of $\uparrow\downarrow$ spin noise
component at $\omega=0$. Introducing some source of an external
spin relaxation, however, leads to a violation of spin
conservation, and amounts in a finite
$S_{LR}^{\uparrow\downarrow}(\omega=0)\ne 0$~\cite{longPRB}.

The fundamental difference between  $\uparrow \uparrow$  and
$\uparrow \downarrow$ correlations shows up even more strikingly
in the spin-conductance (see Fig.~\ref{fig:universal_functions}):
While $G_{LR}^{\uparrow\uparrow}(\omega)$ is dominated by $\tilde
g(\omega)$ and behaves qualitatively the same way as the
conductance through the dot,
$G_{LR}^{\uparrow\downarrow}(\omega)\sim g(\omega)$ exhibits a
resonance at a frequency $\omega \approx 0.5\; T_K $~\footnote{In
the numerical calculations, we define $T_K$ as the half-width of
the composite fermion's spectral function.}. This can be
understood in a simple and intuitive way: The spin conductance
$\uparrow\downarrow$ is generated by  flips of the localized spin.
For $\omega>T_K$, the coupling to the conduction electrons gets
stronger with decreasing $\omega$, and increases the conductance.
At very small energy scales, $\omega\ll T_K$, however, the
impurity spin is quenched, and with the above mechanism being
absent, the $\uparrow\downarrow$ conductance must vanish.


{\bf \em $T \neq 0$, perturbative regime.} Computation of the
finite temperature noise requires care: Usual finite temperature
NRG broadening procedures lead to an unphysical finite linear
coefficient for  $s(\omega)$, conflicting with our exact finite
$T$ result, $s(\omega)\sim \omega^2$. Therefore, for $T\ne 0$,
other methods must be used. For $T\gg T_K$, we  carried out a
systematic expansion in $j$ for the time-dependence of the reduced
density matrix of the spin and the spin current noise using the
formalism of  Refs.~\cite{Konig,Braun}. Details of this involved
calculation shall be published elsewhere~\cite{longPRB}, here we
just outline the main results.

\begin{figure}[t]
  \includegraphics[width=0.95\columnwidth,clip]{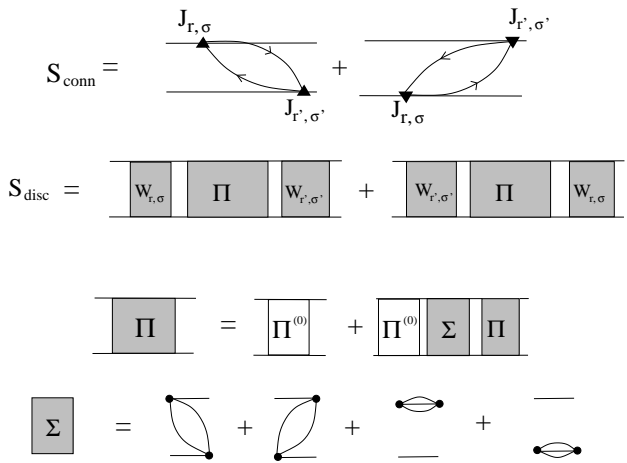}
  \caption{
   Diagrams contributing to the noise.
   The reduced density matrix
   of the spin evolves along the upper and lower
   Keldysh contours. Triangles denote the bare current vertices, dots
   indicate the exchange interaction. Arrows correspond to conduction electron
   propagators. To leading order, the current-current correlation function is
   given by the connected diagrams (top), $S_{\rm conn}(\omega)$,
   and the ``disconnected'' diagrams
   (second line), $S_{\rm disc}(\omega)$,
   with $\Pi$ the propagator describing
   the evolution of the reduced density matrix of the
   spin (third line). The dressed current vertex $W_{r\sigma}$
   is given by diagrams similar to those of the
   self-energy, $\Sigma$ (last line),
   with one of the dots replaced by a triangle.}
  \label{fig:Keldysh}
\end{figure}

\begin{figure}
\includegraphics[width=0.7\columnwidth,clip]{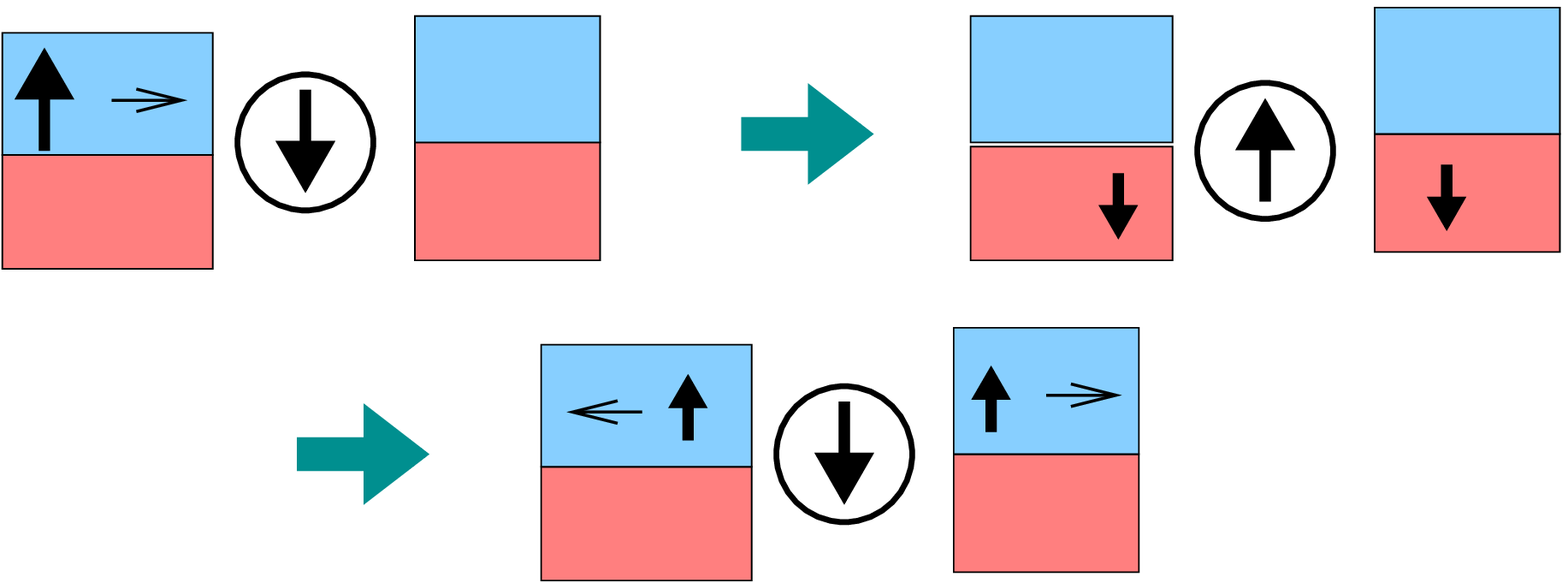}
\null\vskip 0.7cm
\includegraphics[width=0.75\columnwidth,clip]{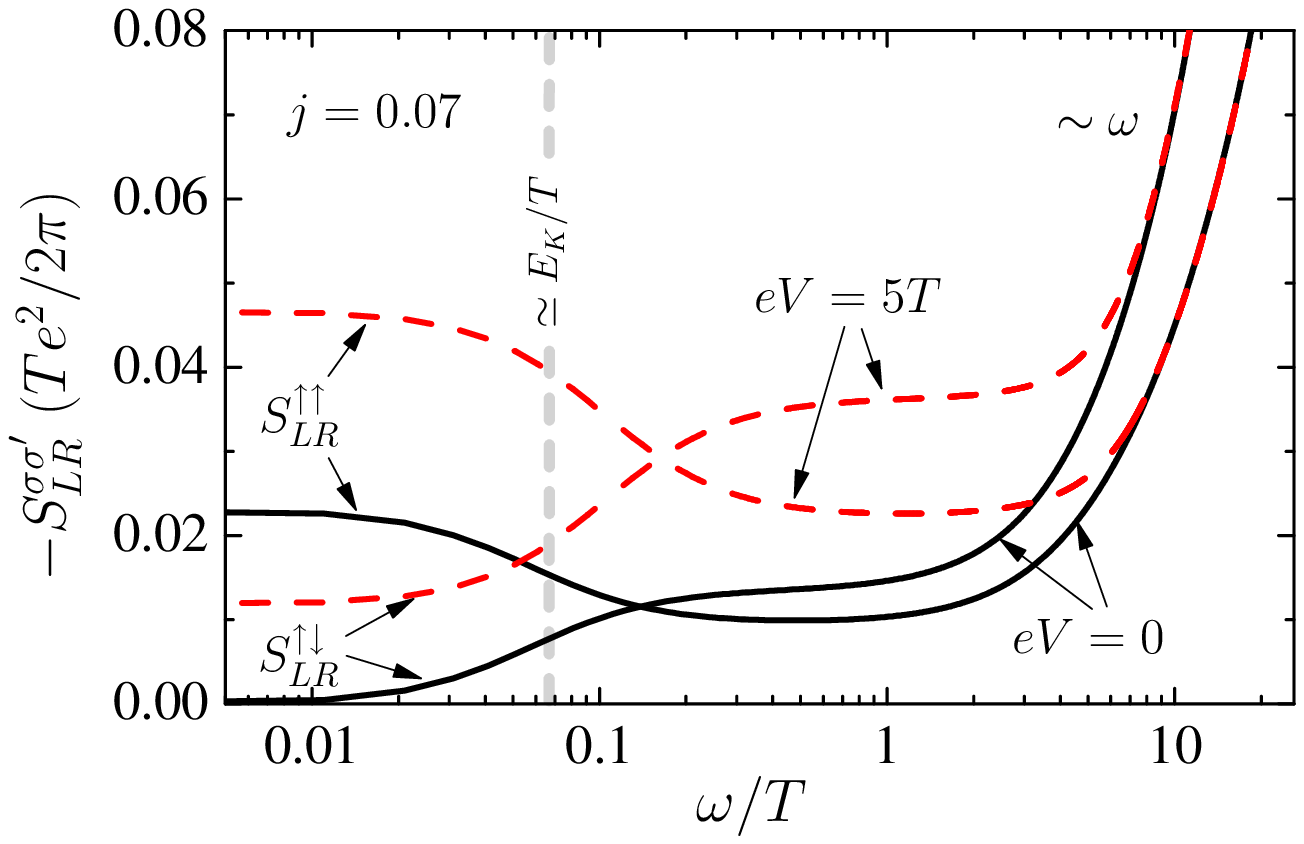}
\caption{\label{fig:noise}
  (color online) Top: Sketch of consecutive spin flip processes.
  Bottom: Equilibrium and non-equilibrium noise spectra
  $S_{LR}^{\s\s'}\left(\omega\right)$ in the perturbative regime,
  $\max\{T,\omega,E_K\}>T_K$, as computed from a diagrammatic
  approach. In the non-equilibrium case a simple current bias was
  assumed, $V_L^\s\equiv V$ and $V_R^\s\equiv 0$, while $j=0.07$
  and $\phi=\pi/2$ in both cases.}
\end{figure}

Naively, to calculate the noise in leading order, one would just
compute the first (connected) noise diagram of
Fig.~\ref{fig:Keldysh}, $S_{\rm conn}(\omega)$. This diagram
accounts for short time current correlations mediated by
electron-hole excitations in the leads, and dominates indeed the
noise at high and intermediate frequencies, $\omega\gtrsim T$. At
small frequencies, however, a resummation of the perturbation
series is necessary, because there the ``disconnected''
contribution, $S_{\rm disc}(\omega)$, turns out to be  of the same
order in $j$ as  $S_{\rm conn}(\omega)$, and becomes also
important: This contribution accounts for correlations between
subsequent {\em incoherent} tunneling processes, generated by the
impurity spin itself. These  correlations are due to the mere fact
that a spin flip process where a conduction electron's spin is
flipped from up do down, $\uparrow\to\downarrow$, must be followed
by a process  $\downarrow\to\uparrow$. To account for them, one
needs  to solve a Dyson equation for the propagator $\Pi$ of the
reduced density matrix of the spin, as sketched in
Fig.~\ref{fig:Keldysh}. In this approach, spin relaxation is
characterized by the relaxation rate, $\Gamma(\omega)$, appearing
in the  self-energy $\Sigma$ of the propagator $\Pi$~\cite{Konig},
\be
\Gamma(\omega) = T\;\sum_{r,r'} j^2 v_r^2 v_{r'}^2 \; \hat
L\left ( \frac \omega T,\frac{\mu_{r}-\mu_{r'}}T\right )\;,
\ee
where ${\rm Re}\;\hat L(x,y) = \pi \;y + \pi [x \;{\rm sh}(x)-y
\;{\rm sh}(y)]/[{\rm ch}(x)-{\rm ch}(y)]$, and ${\rm Im}\;\hat
L(x,y) = \frac 1 \pi \int dx' {\rm Re}\;\hat L(x',y)/(x-x')$. In
the $\omega\to 0 $ limit, $\Gamma(\omega)$ can be identified as
the Korringa relaxation rate, $E_K\equiv  \Gamma(0)/2$, of the
impurity spin, which for a simple voltage-biased quantum dot reads
\begin{equation}
E_K  =  \pi j^2 T \left[  \frac{1+\cos\phi}{2}+
\frac{1-\cos\phi}{2}\; \frac{eV}{2T}\coth\frac{eV}{2T} \right] \;.
\nonumber
\end{equation}
In the voltage-biased case, we can express the left-right
component of the spin current noise in a compact form
\bea \lefteqn{ S^{\s\s'}_{LR}(\omega) = - \frac{e^2}{2\pi} \;{\rm
Re} \Biggl[\frac{\s \s'}{16}\;\frac{\Gamma^2(\omega) -
R^2(\omega)} {-i\omega + \Gamma(\omega)/2}} \label{eq:S_pt}
\\&&
+\frac{3-\s\s'}{32} T j^2  \sin^2(\phi)\; \Bigl[ \hat L \left (
\frac{\omega}{T},\frac{V}{T}\right ) +  \hat L \left
(\frac{\omega}{T},\frac{-V}{T}\right ) \Bigr]\Biggr] \;, \nonumber
\eea
with $R(\omega)= j^2 \cos(\phi) \;\hat L(\omega/T,0)$. The
symmetrized noise is shown in Fig.~\ref{fig:noise}: At high
frequencies, $\omega\gg E_K$, the noise is dominated by the result
of simple-minded perturbation theory, corresponding to the second
line of Eq.~\eqref{eq:S_pt}. This part of the correlation function
describes {\em short-time} correlations within a single tunneling
process, generated by the dynamics of  electron-hole excitations
in the leads. However, at time scales $t\sim 1/\omega>1/E_K$,
consecutive incoherent tunneling processes start to correlate by
the constraint mentioned before. These correlations are captured
by the first term in Eq.~\eqref{eq:S_pt}, coming from the
``disconnected'' part of the noise (see Fig.~\ref{fig:Keldysh}).
As a consequence, for $\omega < E_K$, a large dip appears in the
noise component $S_{LR}^{\uparrow\downarrow}$, while a bump
emerges in $S_{LR}^{\uparrow\uparrow}$. For zero-bias, $V=0$, we
find that $S_{LR}^{\uparrow\downarrow}(\omega=0,V=0)=0$, in
agreement with the fluctuation-dissipation theorem and the
observation that the linear spin conductance between spin up and
spin down electrons must vanish. This has a simple physical
explanation: A spin-$\uparrow$ electron injected from the left can
give rise to a spin-$\downarrow$ outgoing electron on the right
with a certain probability (see Fig.~\ref{fig:noise}). However,
before such a process occurs again, another spin flip process must
take place, where the dot spin is flipped back. In equilibrium,
this second process (on the average) removes {\em exactly} the same
amount of $\downarrow$ spin from the right lead as injected in the
first process. Therefore, no equilibrium $\uparrow\downarrow$ {\it
dc} spin conductance is possible.

Remarkably, the above correlations only show up in the spin
current noise, and cancel out in the charge current noise,
$S_{LR}\equiv \sum_{\s,\s'} S_{LR}^{\s \s'} $. Being the result of
rather classical correlations between subsequent incoherent
processes, these low frequency features can also be captured by a
much simpler rate equation approach (see
Refs.~\cite{KindermannPRB2005,longPRB}), which, however, is unable
to account for the high frequency part of the noise at $\omega>T$.

Although the above results are perturbative in $j$, they carry
over to the whole regime $\max\{T,\omega,E_K\}> T_K$ with the
small modification  that $j$ must be replaced by the renormalized
coupling $j\to j(T,\omega,eV)\approx 1/\ln(\max\{T,\omega,E_K\}/T_K)$.


{\bf \emph{Fermi liquid regime.}} In the Fermi liquid regime,
$\omega,T,eV\ll T_K$, one can compute the spin current
correlations by describing the dot in terms of  scattering states
that interact at the impurity site. This is a rather cumbersome
approach for finite frequencies. However, observing that
correlations between spin up and down electrons are generated only
through the residual electron-electron interaction, simple phase
space arguments immediately give that the $T=0$ shot noise is just
given by $S_{LR}^{\uparrow\downarrow}(V) =(e^2/2\pi)\; \gamma
\;\sin^2(\phi) (eV)^3/T_K^2$, while for equilibrium we recover the
numerically observed result, $S_{LR}^{\uparrow\downarrow}(\omega)
= (e^2/2\pi)\; \alpha \; \sin^2(\phi) |\omega|^3/T_K^2$, with  $\gamma$
and $\alpha$ two universal numbers. The discussion of the finite
temperature and finite frequency noise and the precise
determination of these universal constants is  very complicated,
and shall be considered in a future publication.


{\bf \emph{ Conclusions}.} Analyzing the full frequency dependence
of the spin current noise through a quantum dot in the Kondo
regime we found that $\uparrow\downarrow$ correlations are
strongly suppressed at frequencies  below the Kondo temperature
and below the Korringa relaxation rate as compared to
$\uparrow\uparrow$ correlations due to overall spin conservation.
In the $\uparrow\downarrow$ conductance a resonance is predicted
at $\omega\sim T_K$. Observing these striking features is within
reach with present-day noise measurement techniques.

%
We would like to thank Laci Borda for useful suggestions. 
This research has been supported by Hungarian grants OTKA Nos.
NF061726, K73361, and Romanian grant CNCSIS PN II ID-672/2009, and the EU 
GEOMDISS project. I.W. acknowledges support from the Foundation for
Polish Science and the Ministry of Science and Higher Education
through a research project in years 2008-2010.


\end{document}